# Time-resolved Raman spectroscopy of optical phonons in graphite: Phonon Anharmonic Coupling and Anomalous Stiffening

Hugen Yan, Daohua Song, Kin Fai Mak, Ioannis Chatzakis[1],

Janina Maultzsch[2], and Tony F. Heinz*

*Departments of Physics and Electrical Engineering, Columbia University*
*New York, NY 10027*

[1]*Current affiliation*: Department of Physics, Kansas State University, Manhattan, KS 66506

[2]*Current affiliation*: Institut für Festkörperphysik, Technische Universität Berlin, Germany

**Abstract:**

Time-resolved Raman spectroscopy has been applied to probe the anharmonic coupling and electron-phonon interaction of optical phonons in graphite. From the decay of the transient anti-Stokes scattering of the G-mode following ultrafast excitation, we measured a lifetime of 2.2 ± 0.1ps for zone-center optical phonons. We also observed a transient stiffening of G-mode phonons, an effect attributed to the reduction of the electron-phonon coupling for high electronic temperatures.

Because of its intimate relation with nanotubes and graphene, graphite has recently attracted renewed attention. The interactions between its fundamental excitations -- electron-electron, electron-phonon and phonon-phonon -- play crucial roles in determining the basic physical properties of graphite. Ultrafast pump-probe spectroscopy, by permitting us to achieve non-equilibrium conditions, provides a powerful probe of these interactions [1-5]. Analysis of the dynamics indicates that optical phonons play an important or even dominant role in the relaxation of the excited system [3]. Further, an understanding of phonon dynamics and phonon-phonon interactions is crucial in defining high-field transport properties of graphitic materials [6, 7].

In this Letter, we present measurements of the *ultrafast dynamics of phonons* in graphite. Through application of femtosecond time-resolved Raman scattering [8, 9], we trace the generation of non-equilibrium optical phonons by carrier cooling, their subsequent interaction with electronic excitations, and their decay through anharmonic coupling to lower-energy phonons. The experimental approach permits a direct determination of the absolute phonon mode population and its temporal evolution following femtosecond laser excitation. It complements a recent independent study by Ishioka *et al.* [10] in which ultrafast reflectivity measurements were used to trace the dynamics of coherent phonons in graphite. In our investigation we have established a broad understanding of the role of phonons in the ultrafast dynamics in graphite. We find that: (1) photo-excited carriers transfer most of their energy to a set of strongly-coupled optical phonons (SCOPs), including the zone-center (G-mode) phonons, within a few hundred femtoseconds. This produces a significant non-equilibrium phonon population. The electronic excitations retain only a minor fraction of the initial excitation energy. (2) The optical phonons cool with a time constant of 2.2 ± 0.1 ps. Energy flows from the SCOPs to lower-energy phonons by anharmonic coupling. This process also cools the coupled electronic excitations in graphite. (3) In the transient regime, the non-equilibrium G-mode phonons stiffen, increasing in energy by as much as 10 cm$^{-1}$ compared with the unexcited system. This unusual behavior reflects a decrease in the electron-phonon coupling at high electronic temperatures and the corresponding reduction in the renormalization of the phonon energy.

In our experiment the graphite sample was excited by laser pulses with a duration of ~100 fs and a wavelength of 800 nm as provided by a regeneratively amplified, modelocked Ti:sapphire laser operating at a 1-kHz repetition rate. The response of the sample was probed by Raman scattering induced by 400-nm pulses with duration of ~150 fs. In our measurements, the typical absorbed pump fluence was 2.0 J/m$^2$, while the probe fluence was < 0.2 J/m$^2$. The scattered probe light was detected by a CCD (charge-coupled device) array after dispersion in a single-pass optical spectrometer. The samples were commercial (Advanced Ceramics) highly-oriented pyrolytic graphite (HOPG) with a mosaic spread of 0.8°. The sample surfaces were freshly cleaved along the basal plane prior to the measurements.

Figure 1 displays the G-mode Raman scattering spectra of graphite measured with the ultrashort probe pulses. The anti-Stokes (main figure) and Stokes (inset) spectra are shown both in the presence and absence of simultaneous pump excitation. Both the anti-Stokes and Stokes Raman intensities are enhanced by the pump pulses, with the anti-Stokes Raman signal increasing by more than an order of magnitude for the absorbed pump fluence of 2.0 J/m$^2$. The spectra also reveal another effect: The phonon frequency is blue shifted under pump excitation.

In Fig. 2 we present measurements of the Raman scattering as a function of the probe delay time. Fig. 2(a) displays the temporal evolution of the anti-Stokes Raman signal. The corresponding central frequency of the G-mode phonons is shown in Fig 2(b). A time-dependent frequency shift can be seen. We will return to a discussion of this aspect of the phonon dynamics later in the paper.

The strength of this anti-Stokes scattering signal in Fig. 2(a) is proportional to the time-dependent mode population $n_G$ of the zone-center optical phonons. The dynamics of this response did not show meaningful differences as the pump fluence was varied (from 1 – 5 J/m$^2$) and results are presented for a pump fluence of 2.0 J/m$^2$. We established the *absolute mode population* $n_G$ by analysis of the enhancement of the Stokes scattering (inset of Fig. 1). Since the Stokes intensity scales as $1 + n_G$ and $n_G << 1$ for the unpumped sample, the fractional increase in the Stokes scattering intensity yields $n_G$ directly. (We established the consistency of the value of $n_G$ with that deduced from a comparison of the antiStokes and Stokes scattering intensities, calibrated using an unpumped sample at room temperature.) In our experiment we used this method to calibrate $n_G$ at zero-time delay, where $n_G$ is relatively large. We then obtained $n_G$ for other delay times by comparing the anti-Stokes scattering intensities with that at zero time delay. In Fig. 2(c) we also display the G-mode phonon population $n_G$ in terms of an effective temperature *T*. This temperature is defined in terms of the mode population using the Bose-Einstein distribution of $n_G = 1/[\exp(\hbar\omega_G/kT)-1]$, where $\hbar\omega_G$ is the phonon energy.

The temporal evolution of the G-mode phonon population can be fit to simple exponential decay, after including the instrumental response [Fig. 2(a)]. We deduce thereby a lifetime of τ = 2.2 ± 0.1 ps for the G-mode phonons, where the indicated error is based on the analysis of several separate measurements. By examining the rising edge of the phonon population, we also obtain an upper bound for the build up of the G-mode phonon population of ~300 fs.

The rapid build-up of G-mode phonons from the initial electronic excitation of the graphite by a short laser pulse is not unexpected. Theoretical studies have predicted highly efficient coupling of the hot carriers to Γ-point (G-mode) optical phonons, as well as those near the K-point [11, 12]. Further, experimental studies of graphite electron dynamics by time-resolved THz and photoemission spectroscopy [2, 3] indicate that cooling occurs on the sub-picosecond time scale, with the emission of G-mode optical phonons expected to be one of the dominant channels. The recent observation of coherent G-mode phonons under excitation with ultrafast laser pulses [10, 13] implies the existence of a nearly instantaneous contribution (within the 20-fs phonon period) to the generation of G-mode phonons upon optical excitation of graphitic compounds.

A striking feature of our data is the value of mode population $n_G$ that is achieved by optical pumping. We measure, for an absorbed fluence 2.0 J/m$^2$, a G-mode population of $n_G = 0.32$, corresponding to an effective temperature of 1600 K. This result permits us to draw some important conclusions about the transient distribution of energy in graphite. For the optical penetration depth of 800-nm radiation in graphite of 15 nm [14], our absorbed fluence of 2.0 J/m$^2$ implies a density of deposited energy near the surface of 5.2 meV per C atom. In full thermal equilibrium, this only leads to a ~50 K temperature increase [15]. The G-mode phonons, which reach an effective temperature of 1600 K, clearly exhibit strongly non-equilibrium behavior. From an energy balance argument, we see that only a small fraction of the optical phonons (~ 8% of the Brillouin zone) can be excited to this extent. This behavior is consistent with a picture in which the photoexcited electrons and holes, through intravalley and intervalley scattering, can only produce phonons close to Γ- and K-points because of the requirements of momentum conservation[16]. Finally because of the strong electron-phonon coupling and rapid electron-electron scattering [2, 3, 11, 12], we assume that the electronic excitations are described by the G-mode temperature *T*. From the electronic specific heat for graphite [15], we then find that the additional energy in the electronic excitations at 1600 K amounts only to 0.47 meV per C atom. We have thus shown that under our excitation conditions over 90% of the energy resides in the SCOPs.

In the analysis above (and in our discussion of the temporal evolution of the phonon population that follows), we assume that spatial redistribution of the deposited energy can be neglected.

Given the spot size of ~ 1 mm in our measurement, lateral spreading of excitation, even in the presence of the much higher in-plane diffusion rates, occurs only on the microsecond time scale. Perpendicular to the surface, however, there is a much strong gradient in the excitation density arising from strong optical absorption of the pump beam. Despite this gradient, as evidenced by comparison of the ultrafast transient reflection and transmission measurements by Seibert *et al.* [1], the weak coupling between the graphene layers of the graphite structure implies that vertical energy migration remains insignificant on the time scale of several picoseconds.

From the discussion above, we see that the initial condition for the decay of the optical phonon population is one in which a subset of SCOPs is at a high temperature, but with most of the phonon modes exhibiting only their initial room-temperature thermal excitation. We then can attribute the observed 2.2-ps lifetime of the G-mode population to the decay of these phonons into lower-energy phonons through anharmonic coupling. More precisely, within the picture of a subsystem of SCOPs comprised of both $\Gamma$- and K-point phonons held in equilibrium with the electronic excitations, the measured phonon decay rate should be regarded as the average anharmonic decay rate for all of the SCOPs, including both those near the $\Gamma$- and K-points. In addition to the role of electronic excitations in mediating energy exchange between different SCOPs, we should consider their effect on the storage and dissipation of energy. As argued above, the non-equilibrium SCOPs account for nearly all of the deposited energy, so that electronic excitations do not provide a significant reservoir of energy that could alter phonon decay rates. Similarly, direct excitation of lower energy acoustic phonons by electronic excitations is not expected to be important, since electron energy loss in each accessible scattering event will be very modest.

*Ab-initio* calculations have examined possible routes for anharmonic decay of SCOPs in graphene [17]. The G-mode phonons decay principally into two longitudinal acoustic phonons or one longitudinal and one transverse acoustic phonon. The anharmonic lifetime for the G-mode phonons at room temperature was calculated to be 3 ps, in reasonable agreement with our experimental result. It is also interesting to compare our experimental finding for the G-mode lifetime in graphite to that reported for (6,5)-index semiconducting nanotubes [18]. The anharmonic lifetime in the latter case was 1.1 ps, just half of the value of 2.2 ps reported here for graphite. Two important distinctions may be drawn between the different material systems. The first is that the measured optical phonon lifetime in graphite may be influenced by the behavior of K-point optical phonons, as discussed above. Calculations suggest that K-point optical phonons have weaker anharmonic coupling and, hence, longer lifetimes [17]. A second factor is the existence in nanotubes of additional lower-energy phonon modes, such as the radial breathing mode, that exhibit strong coupling to G-mode phonons [13] and may open up new anharmonic decay pathways.

As a final remark about the anharmonic lifetime, we compare our pump-probe time-domain results with data from conventional frequency-domain Raman measurements. The 2.2 ps anharmonic lifetime of G-mode phonons implies a 2.4 cm$^{-1}$ (FWHM) contribution to the width of the Raman feature. In addition, however, one expects a contribution to the Raman line width from the strong coupling of G-mode phonons to electron-hole pairs.  From measurements of the effect of electrostatic gating, the strength of this coupling for graphene has been deduced to be 8.5 cm$^{-1}$ [19].  Assuming a similar contribution for graphite, the total lifetime broadening of the Raman line width would be ~11 cm$^{-1}$. The measured G-mode width is ~13 cm$^{-1}$ [20-22].  The modest residual broadening may arise from inhomogeneous broadening and pure dephasing contributions.

In addition to the changes in the strength of the Raman scattering, we also observed a clear time dependence in G-mode frequency [Fig. 2(b)]. For most of materials, including graphite [22] and graphene layers [23], phonon frequencies *decrease* under equilibrium heating.  Similarly phonon softening has also been observed for solids with strong electronic excitation under nonequilibrium excitation conditions [24].  Surprisingly, however, in our transient measurements of graphite, we observe an *increase* in the G-mode frequency upon excitation.  A similar effect was also recently reported by Ishioka et al. [10] for coherently generated phonons.

To understand the origin of the transient stiffening of the G-mode phonons, we first recall the influence of electron-phonon coupling on the phonon frequency in graphitic materials. As in graphene layers and metallic nanotubes [19, 25-29], the phonon dispersion in graphite exhibits Kohn anomalies for the highest energy branches at $\Gamma$-and K-points [11], with the strong electronic coupling of the phonons leading to a reduction in the phonon energy. We attribute the observed stiffening of the G-mode phonons to a reduction of this electronic coupling for the high transient electronic temperatures[30]. Such behavior is closely related to recent experimental demonstrations of the changes in the frequency (and lifetime) of the G-mode phonons induced in graphene [19, 25, 29] and metallic nanotubes [26-28] by electrically tuning the Fermi level. When the Fermi level is near the Dirac point, G-mode phonons decay efficiently into electron-hole pairs. However, if the Fermi level is moved away from the Dirac point, this process can be switched off once the required electron and hole states are no longer available. This electronic decoupling leads to an increase in the phonon energy. In our experiment, the high transient electronic temperature acts in the same way.

To be more quantitative, we consider the self-energy of G-mode phonons as a function of the electronic temperature $T$. For simplicity, we approximate the graphite band structure by that of graphene, *i.e.*, by a two-dimensional linear dispersion relation. The formalism introduced by Ando [31] then provides an explicit expression for the phonon self-energy:

$$\Pi = \lim_{\delta \to 0}\left\{-\lambda \int_0^{+\infty} \varepsilon d\varepsilon [f_T(-\varepsilon) - f_T(\varepsilon)] \times \left(\frac{1}{\hbar\omega_G + 2\varepsilon + i\delta} - \frac{1}{\hbar\omega_G - 2\varepsilon + i\delta}\right)\right\},$$

where $\lambda$ denotes the dimensionless electron-phonon coupling strength and $f_T(\varepsilon)$ the Fermi-Dirac distribution function. In this analysis of the phonon self-energy, we neglect any change in the anharmonic contribution. This is a reasonable assumption since we expect only a modest temperature rise for the lower energy phonons into which the optical phonons can decay anharmonically. Fig. 3 shows the predicted shift in the G-mode energy (from Re $\Pi$) as a function of temperature. The results are compared with the temperature-dependent shifts obtained from the experimental data in Figs. 2(b) and (c). Only delay times > 0.2 ps are included in the analysis since it requires equilibration between the electronic excitations and the SCOPs. A good fit to theory is obtained for an electron-phonon coupling of $\lambda = 0.0064$, corresponding to an electron-phonon deformation potential of $D = 12.1$ eV/Å[19, 31]. This value is quite close to that inferred from measurements of electrically gated graphene samples [19].

In summary we have applied time-resolved Raman scattering to probe the ultrafast dynamics and interactions of optical phonons in graphite. The measurements provided quantitative information on the zone-center phonons, including their population and energy content, anharmonic lifetimes, and electron-phonon interactions. The results show the crucial role of these optical phonons in the ultrafast electron dynamics in graphitic materials and support the importance of a subset of strongly coupled optical phonons that readily thermalize with electronic excitations in the system. While this general picture is expected to apply for various forms of $sp^2$ – hybridized carbon, the origin of differences in phonon dynamics between different forms of graphite carbon remains an open question. We thank Dr. Yi Rao for experimental assistance and acknowledge support from the Nanoscale Science and Engineering Initiative of the NSF under grants CHE-0117752 and ECS-05-07111, the New York State Office of Science, Technology, and Academic Research (NYSTAR), the Chemical Sciences, Geosciences, and Biosciences Division, Office of Basic Sciences, Office of Science, U. S. Department of Energy the Office of Basic Energy Sciences, and from the Alexander-von-Humboldt Foundation for JM.

**Figures and Figure Captions:**

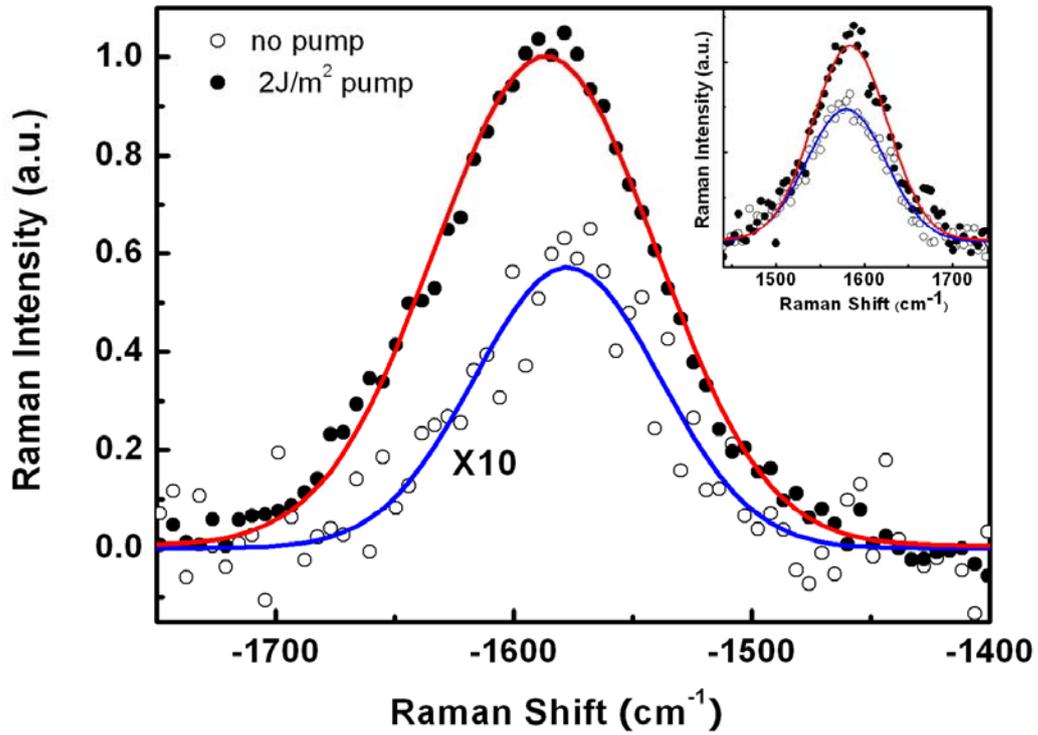

Fig. 1. Raman spectra of the G-mode phonons in graphite with and without simultaneous pump excitation. The main figure shows the anti-Stokes scattering; the inset, the Stokes scattering response. The solid lines are fits to a Gaussian line shape, with a width determined by the probe laser spectrum.

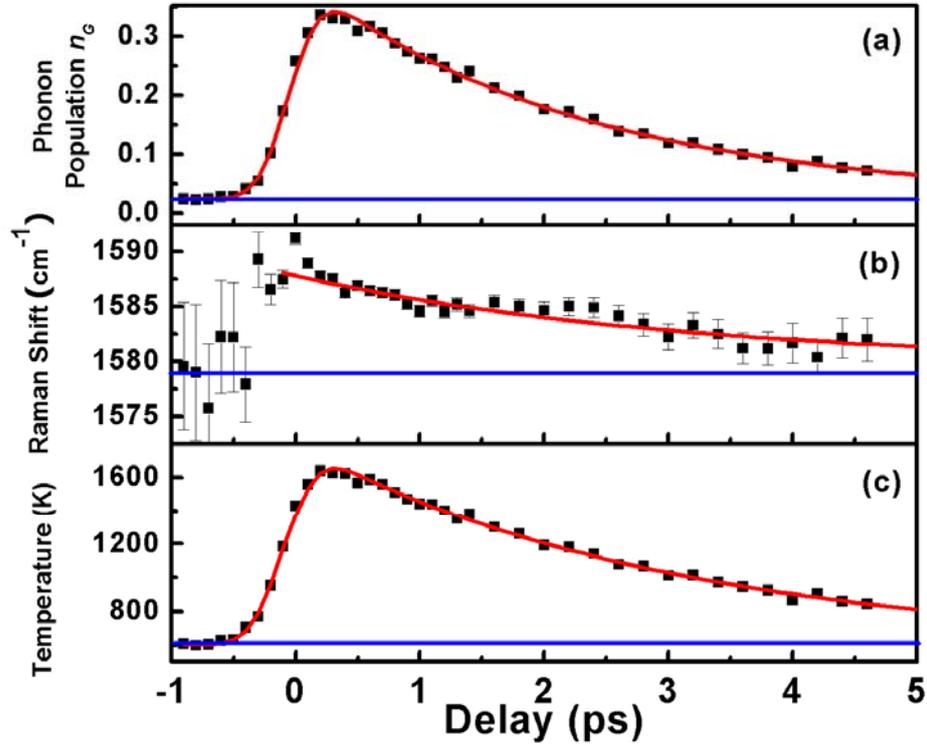

Fig. 2. Temporal dynamics of the G-mode phonons as a function of delay time following pump excitation. (a) Experimental anti-Stokes Raman intensity, which is proportional to the phonon mode population $n_G$. The solid line represents exponential decay with a time constant of 2.2 ps convoluted with the instrumental response. The absolute calibration of $n_G$, as discussed in the text, is based on the enhancement of the Stokes Raman signal. (b) Measured shifts of the G-mode frequency. The error bars are derived from fits like those in Fig. 1. The solid line is based on the model of the temperature dependence of the self-energy described in the text. (c) Temperature of the G-mode phonons inferred from the mode population in (a). At negative delay, the temperature is higher than room temperature. This increased temperature arises from self-pumping by the probe pulse.

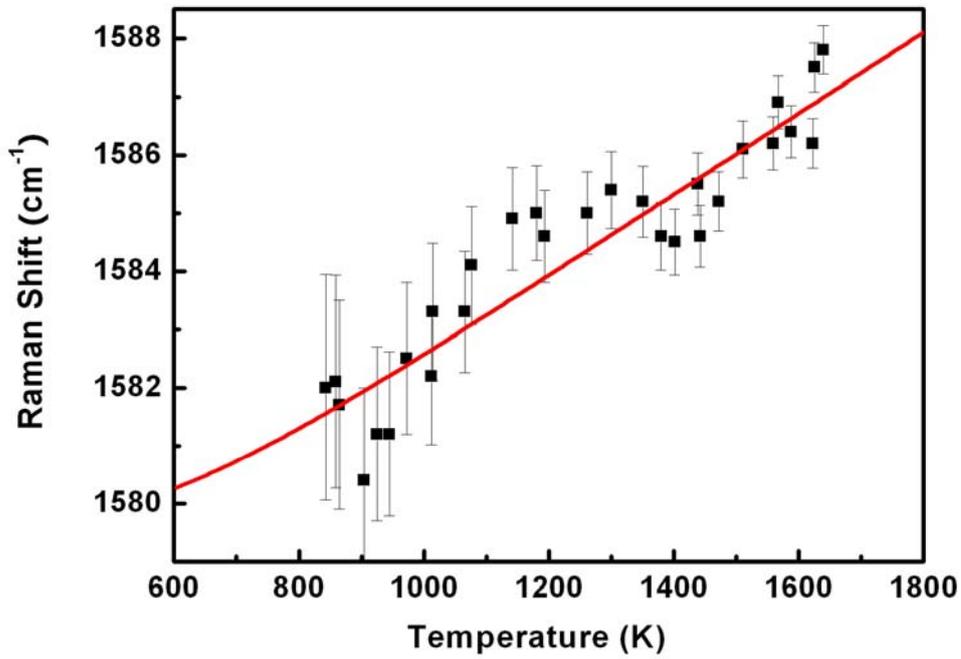

Fig. 3. Raman shift for different electronic temperatures. The data points are obtained from Figs. 2(b) and (c) for delay times > 0.2 ps. The solid line is the result of the calculation of the phonon self-energy, as presented in the text, for an electron-phonon coupling strength of $\lambda = 0.0064$. The apparent oscillatory behavior of the data points could not be reproduced in other measurements and is attributed to experimental uncertainty.